# An analysis and solution of ill-conditioning in physics-informed neural networks

Wenbo Cao [a,b], Weiwei Zhang [a,b,*]

[a] *School of Aeronautics, Northwestern Polytechnical University, Xi'an 710072, China.*
[b] *International Joint Institute of Artificial Intelligence on Fluid Mechanics, Northwestern Polytechnical University, Xi'an, 710072, China*

**Abstract.** Physics-informed neural networks (PINNs) have recently emerged as a novel and popular approach for solving forward and inverse problems involving partial differential equations (PDEs). However, achieving stable training and obtaining correct results remain a challenge in many cases, often attributed to the ill-conditioning of PINNs. Nonetheless, further analysis is still lacking, severely limiting the progress and applications of PINNs in complex engineering problems. Drawing inspiration from the ill-conditioning analysis in traditional numerical methods, we establish a connection between the ill-conditioning of PINNs and the ill-conditioning of the Jacobian matrix of the PDE system. Specifically, for any given PDE system, we construct its controlled system. This controlled system allows for adjustment of the condition number of the Jacobian matrix while retaining the same solution as the original system. Our numerical findings suggest that the ill-conditioning observed in PINNs predominantly stems from the Jacobian matrix. As the condition number of the Jacobian matrix decreases, PINNs exhibit faster convergence rates and higher accuracy. Building upon this understanding and the natural extension of controlled systems, we present a general approach to mitigate the ill-conditioning of PINNs, leading to successful simulations of the three-dimensional flow around the M6 wing at a Reynolds number of 5,000. To the best of our knowledge, this is the first time that PINNs have been successful in simulating such complex systems, offering a promising new technique for addressing industrial complexity problems. Our findings also offer valuable insights guiding the future development of PINNs.

**Keywords.** PINNs, ill-conditioning, controlled system, Jacobian matrix, condition number.

## 1 Introduction

Recently emerged methods for solving PDEs through neural network optimization, such as physics-informed neural networks (PINNs) [1], deep Ritz method [2], and deep Galerkin method [3] have been widely used to solve forward and inverse problems involving PDEs. By minimizing the loss of PDE residuals, boundary conditions and initial conditions simultaneously, the solution can be straightforwardly obtained without mesh, spatial discretization, and complicated program. With the significant progress in deep learning and computation capability, a variety of PINN-like methods have been proposed in the past few years, and have achieved remarkable results across a range of problems in computational science and engineering [4-7]. As a representative optimization-based PDE solver, PINNs offer a natural method for solving PDE-constrained optimization problems, yielding a lot of valuable research outcomes, including flow visualization technology [8-10], optimal control [11-13] and inverse design for topology optimization [14]. In addition, PINN-like methods have also achieved remarkable results in solving parametric problems [3, 15-18].

Despite the potential for a wide range of physical phenomena and applications, training PINN models still encounters challenges in many complex problems [4]. A typical issue is that many fluid-related applications are limited to relatively low Reynolds numbers [15, 19, 20]. Some studies [21, 22] attribute the ill-conditioning of PINNs to the unbalanced loss between PDE residual and boundary condition residual, proposing to balance the weights of loss components during training. However, this only partially explains the ill-conditioning of PINNs, as challenges persist even for enforced boundary conditions, as indicated in [19]. Some researchers further attribute the ill-conditioning of PINNs to PDE-based soft constraint [23]. Nevertheless, further analysis of the ill-conditioning of PINNs is still lacking, severely constraining the progress of PINNs in addressing complex engineering problems and undermining confidence in this emerging technology. In this study, we explore the ill-conditioning of PINNs starting from the ill-conditioning in traditional numerical methods.

We consider a dynamic system represented as:

$$\dot{q} = f(q) \qquad (1)$$

where the dot expresses the time derivative, $q$ represents the solution of the dynamic system, and $f$ encompasses both the PDE operator and the boundary condition operator. We focus on obtaining the steady solution $q_s$ of the dynamic system, which satisfies $f(q_s) = 0$.

In traditional grid-based numerical methods, the initial step is to discretize the system on a given mesh, yielding a large discrete system

$$\boldsymbol{f}(\boldsymbol{q}) = 0 \qquad (2)$$

where $\boldsymbol{q} \in \mathbb{R}^N$ represents the set of state variables describing the solution at each spatial location of the mesh in the domain $\Omega$, and $\boldsymbol{f}:\Omega \in \mathbb{R}^N \to \mathbb{R}^N$ is differentiable and represents the discrete residuals. Subsequently, addressing this discrete system often involves solving a linear system $A\boldsymbol{q} = \boldsymbol{b}$. In numerous numerical methods, such as those based on the Newton's method, matrix $A$ is the Jacobian matrix $J \in \mathbb{R}^{N \times N}$ corresponds to the linearization of the discrete residual $\boldsymbol{f}$ around $\boldsymbol{q}_0$:

$$J(\boldsymbol{q}_0) = \frac{\partial \boldsymbol{f}}{\partial \boldsymbol{q}}\bigg|_{\boldsymbol{q}=\boldsymbol{q}_0} \tag{3}$$

The Jacobian matrix $J$ is a large sparse matrix that is crucial to the dynamic system. The eigenvalues of $J(\boldsymbol{q}_s)$ characterize the stability of the system. Pursuing a steady solution typically indicates that the system is stable, signifying that all eigenvalues of $J(\boldsymbol{q}_s)$ have a real part less than zero. The condition number of $J(\boldsymbol{q}_s)$ serves as an indicator of the system's ill-conditioning. The convergence speed of many iterative methods is based on the spectral properties of the matrices, and hence ill-conditioned systems can converge slowly. Therefore, in traditional numerical methods, countless preconditioning techniques have been developed over the years to alleviate ill-conditioning and expedite convergence. Computational experience accumulated in the past couple of decades indicates that a good preconditioner holds the key for an effective iterative solver.

Inspired by the ill-conditioned analysis of traditional numerical methods, a reasonable conjecture is whether the ill-conditioning of PINNs also originates from the ill-conditioning of the Jacobian matrix (or the Fréchet derivative in the infinite-dimensional situation, denoted as $D_q f$). Despite being fundamentally an optimization problem, we seek to relate the convergence of PINNs to the system's Jacobian matrix rather than the Hessian matrix of the loss function. This is done to decouple the ill-conditioning of PINNs from neural network, as it is widely acknowledged that the powerful capability of neural networks to approximate nonlinear functions has been extensively validated. Attributing the ill-conditioning of PINNs to complex nonlinear equations rather than neural networks helps to provide a clearer understanding of PINNs' ill-conditioning and establishes new solutions.

The main contributions of our work can be summarized as follows:

1. We propose a controlled system with an adjustable condition number of the Jacobian matrix to visualize the correlation between the ill-conditioning of the Jacobian matrix and that of PINNs.

2. Based on the analysis of ill-conditioning and a natural extension of controlled

systems, we propose a general solution to mitigate the ill-conditioning of PINNs.

3. We successfully solved the flow around the 3-dimensional M6 wing at a Reynolds number of 5,000, which represents nearly the maximum Reynolds number allowable for incompressible laminar flow over this shape.

The remainder of the paper is organized as follows. Section 2 provides a brief introduction to PINNs and reformulates its loss function. In Section 3, we construct a controlled system to visualize the correlation between the ill-conditioning of the Jacobian matrix and that of PINNs. Section 4 proposes an approach for mitigating the ill-conditioning of PINNs and provides validation. Finally, Section 5 presents concluding remarks and directions for future research.

## 2 Physics-informed neural networks

In this section, we briefly introduce PINNs and reformulate its loss function for subsequent analysis. A typical PINN employs a fully connected deep neural network architecture to represent the solution $q$ of the dynamical system. The network takes the spatial $x \in \Omega$ and temporal $t \in [0, T]$ as the input and outputs the approximate solution $\hat{q}(x, t; \theta)$. The spatial domain typically has 1-, 2- or 3-dimensions in most physical problems, and the temporal domain may be nonexistent for time-independent (steady) problems. The result of PINNs is determined by the network parameters $\theta$, which are optimized with respect to PINNs loss function during the training process.

For example, to obtain the steady solution of the dynamic system represented by Equation (1), we calculate the residual $f(q)$ over a series of $m$ collocation points $D = \{x_i\}_{i=1}^{m}$ by automatic differentiation [24], and then minimize the loss

$$\mathcal{L} = \frac{1}{N} \| f(q(\cdot; \theta)) \|^2 \tag{4}$$

Throughout all cases, we employ a fully connected DNN architecture, equipped with the hyperbolic tangent activation functions (tanh), and trained using the limited-memory Broyden-Fletcher-Goldfarb-Shanno (LBFGS) optimizer.

Note that unlike traditional grid-based numerical methods where boundary conditions are precisely embedded, boundary conditions in PINNs are often treated as soft constraints. Therefore, $f(q)$ encompasses both the PDE residual $g(q)$ and the boundary condition residual $h(q)$, requiring a trade-off between different components through appropriate relative weights $\lambda_{PDE}$ and $\lambda_{BC}$, expressed as follows:

$$f(q) = \begin{bmatrix} \lambda_{PDE} g(q) \\ \lambda_{BC} \sqrt{N_g / N_h} \, h(q) \end{bmatrix} = 0 \tag{5}$$

where $N_g$ and $N_h$ are the dimension of the PDE residual and boundary condition residual respectively. $\sqrt{N_g/N_h}$ is introduced to balance the influence of the number of collocation points, thus ensuring equivalence with the loss function of standard PINNs. For time-dependent problems, $f(q)$ should further include initial condition residual $\lambda_{IC}\sqrt{N_g/N_i}\,i(q)$.

## 3 An analysis of ill-conditioning in PINNs

### 3.1 Controlled system

Inspired by the ill-conditioned analysis of traditional numerical methods, we speculate that the ill-conditioning of PINNs originates from the ill-conditioned Jacobian matrix. Unfortunately, we are unable to obtain the explicit Jacobian matrix in PINNs. This limitation prevents us from establishing a connection between the convergence challenges of PINNs and the ill-conditioning of the Jacobian matrix. To address this issue, we construct a controlled system. The system modifies the eigenvalues of Jacobian matrix by adding a linear forcing term to original system based on control theory. The modified system is written as

$$f_c(q) = f(q) - \gamma(q - q_s) = 0, \gamma > 0 \tag{6}$$

where $q_s$ is the steady solution of the system, and $\gamma$ is the control gain. Obviously, the controlled system shares identical solution as the original system, but all eigenvalues of the controlled system's Jacobian matrix are equal to those of the original system minus gain. For a stable system, the real parts of all eigenvalues of the Jacobian matrix are less than 0. Consequently, as the gain increases, all eigenvalues of the controlled system synchronously deviate in the negative direction away from zero, resulting in the system becoming increasingly well-conditioned.

Therefore, although we cannot explicitly obtain the Jacobian matrix and its condition number, we can mitigate its ill-conditioning by adjusting the gain of the controlled system. As the gain tends to infinity, the original PDE-solving task will degrade into a supervised learning task, a well-known well-conditioned problem. By observing the convergence speed and accuracy of PINNs at different gains, we can establish the relationship between the ill-conditioning of PINNs and the ill-conditioning of the Jacobian matrix.

### 3.2 Two illustrative examples

A. Lid-driven cavity flow

We consider a lid-driven cavity problem, which is a classical benchmark in Computational Fluid Dynamics (CFD). The system is governed by the two-dimensional incompressible Navier-Stokes equations:

$$\begin{aligned}&\boldsymbol{u}\cdot\nabla\boldsymbol{u}+\nabla p-\Delta\boldsymbol{u}/\text{Re}=0\\&\nabla\cdot\boldsymbol{u}=0\\&\boldsymbol{u}=(1,0)\quad\text{in }\Gamma_0\\&\boldsymbol{u}=(0,0)\quad\text{in }\Gamma_1\end{aligned} \qquad (7)$$

where $\boldsymbol{u}=(u,v)$ is velocity vector, $p$ is pressure. The computational domain $\Omega=(0,1)\times(0,1)$ is a two-dimensional square cavity, where $\Gamma_0$ is its top boundary and $\Gamma_1$ is the other three sides. Despite its simple geometry, the driven cavity flow retains a rich fluid flow physics manifested by multiple counter rotating recirculating regions on the corners of the cavity as Re increases [25]. In this example, we choose Re = 2,500. This is the flow regime where standard PINNs fail to solve the problem.

We solve controlled systems with varying gains using a network with 5 hidden layers and 128 neurons per hidden layer. We utilize a 500*500 uniform grid to enforce the PDE residual and boundary condition residual, and evaluate the relative $L_2$ error. The relative $L_2$ error between the predicted value $\hat{\boldsymbol{q}}$ and the reference value $\boldsymbol{q}_{ref}$ is defined as $\|\hat{\boldsymbol{q}}-\boldsymbol{q}_{ref}\|_2/\|\boldsymbol{q}_{ref}\|_2$. The relative weights are $\lambda_{PDE}=2$ and $\lambda_{BC}=1$. To facilitate comparison, we also solved the controlled system using the finite difference method (FDM) combined with the Newton-Krylov iteration. The finite difference method allows us to derive an explicit Jacobian matrix and calculate its condition number for controlled systems with varying gains. We estimate the variation of the condition number of the Jacobian matrix with respect to the gain in PINNs by observing those in FDM. For PINNs, the steady solution within the controlled system is obtained by finite difference method on a very fine mesh (1,000*1,000). In the case of FDM, to derive the explicit Jacobian matrix and calculate its condition number while tracking the convergence history, we simply use a coarse mesh (80*80) and employ its own solution as the steady solution within controlled system.

Figure 1 (a) and (b) respectively depict the convergence history of controlled systems with different gains obtained through PINNs and FDM. We observe that FDM achieves stable linear convergence for any gain. As the gain increases, the condition number of the Jacobian matrix of the controlled system significantly decreases (Table 1), thus yielding a faster convergence rate. PINNs fail to achieve any meaningful results for the original system (i.e., the controlled system with $\gamma=0$), consistent with previous observations [26]. With the increase in gain and the consequent decrease in the condition number of the latent Jacobian matrix, PINNs attain faster convergence, akin to FDM. Despite being constrained by the representation capacity of neural networks and optimization challenges, PINNs typically fail to achieve stable linear convergence. We still observe that, at any given error level, larger gains correspond to greater convergence rates, consistent with the convergence behavior in FDM.

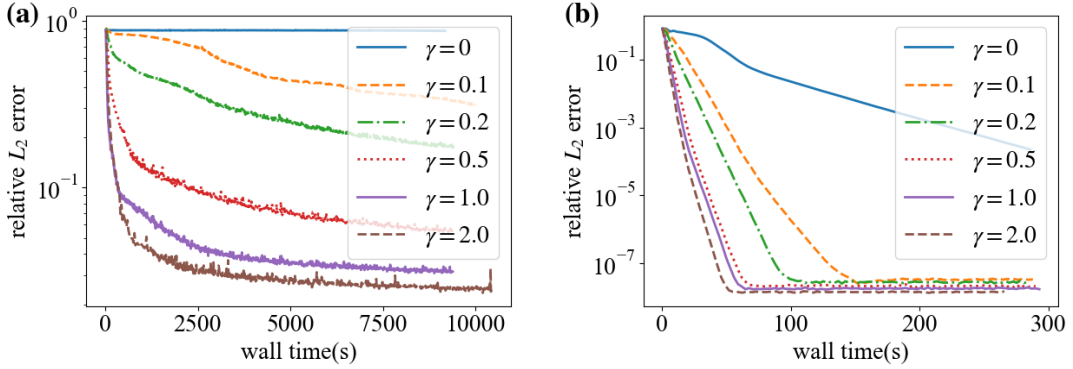

Figure 1. Convergence history of controlled systems with varying positive gains obtained by (a) PINNs and (b) FDM.

Table 1. The condition numbers of the Jacobian matrices of controlled systems with different gains in FDM.

| gain | 0 | 0.1 | 0.2 | 0.5 | 1.0 | 2.0 |
|---|---|---|---|---|---|---|
| condition number | 12542 | 6605 | 4296 | 1982 | 1028 | 502 |
| gain | / | -0.1 | -0.2 | -0.5 | -1.0 | -2.0 |
| condition number | / | $6*10^5$ | $5*10^4$ | $2*10^5$ | $1*10^{16}$ | $5*10^8$ |

To allay a concern that the rapid convergence of the controlled system might be attributed to the introduction of steady solution, we also illustrate the convergence behavior for negative gains, as depicted in Figure 2. Since negative gains may lead to eigenvalues very close to or greater than 0, they typically exacerbate the condition number of the system (Table 1), resulting in unpredictable convergence behavior in FDM (Figure 2(b)). For optimization-based solvers like PINNs, the introduction of steady-state solution in the controlled system often results in rapid error reduction in the initial optimization stages. However, as negative gains fail to improve the condition number of Jacobian matrix, all errors eventually stagnate, precluding stable descent, as shown in Figure 2(a).

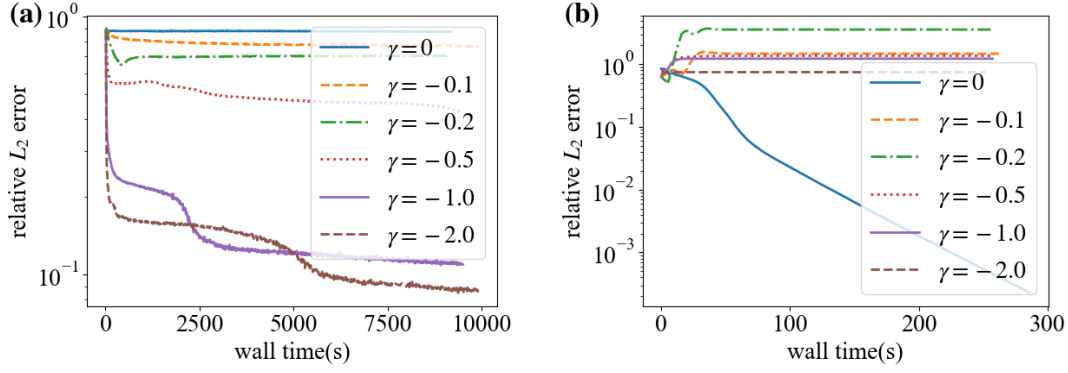

Figure 2. Convergence history of controlled systems with varying negative gains obtained by (a) PINNs and (b) FDM.

B. Allen-Cahn equation

Although we initiate our analysis with a consideration of time-independent dynamical systems, the analysis in Section 3.1 evidently extends to time-dependent problems as well. To further corroborate the applicability of the current perspective to time-dependent problems, we examine the one-dimensional Allen-Cahn equation, a benchmark in PINNs. We solve controlled systems with varying gains using a network with 4 hidden layers and 128 neurons per hidden layer. We utilize a 257*101 uniform grid to enforce the PDE residual and boundary condition residual. The relative weights are $\lambda_{PDE}=1$, $\lambda_{BC}=0.1$ and $\lambda_{IC}=5$.

Figure 3 (a) and (b) respectively illustrate the convergence history of controlled systems with different positive and negative gains obtained through PINNs. Table 2 shows the condition numbers of the Jacobian matrices under different gains in FDM, providing a rough estimation of those in PINNs. We observe that for positive gains, as the gain increases and the potential condition number of the Jacobian matrix decreases, PINNs achieve progressively faster convergence, consistent with the observations in Figure 1. For negative gains, as the system's condition number does not undergo any improvement, despite the controlled systems experiencing rapid initial descent due to the introduction of steady solution, they ultimately fail to exhibit any improvement in convergence behavior compared to the original system.

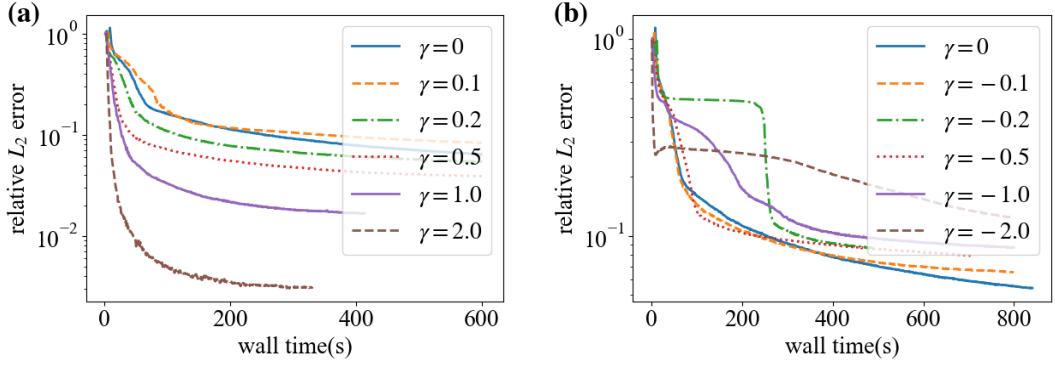

Figure 3. Convergence history of controlled systems with (a) positive gains and (b) negative gains.

Table 2. The condition numbers of the Jacobian matrices of controlled systems with different gains in FDM.

| gain | 0 | 0.1 | 0.2 | 0.5 | 1.0 | 2.0 |
|---|---|---|---|---|---|---|
| condition number | 3448 | 3168 | 2913 | 2271 | 1518 | 716 |
| gain | / | -0.1 | -0.2 | -0.5 | -1.0 | -2.0 |
| condition number | / | 3754 | 4090 | 5302 | 8247 | 20533 |

The numerical examples above suggest that the ill-conditioning of PINNs primarily stems from the ill-conditioning of the Jacobian matrix. Mitigating the ill-conditioning of the Jacobian matrix contributes to stable convergence of PINNs in complex problems.

## 4 A solution of ill-conditioning in PINNs

4.1 Improved time-stepping-oriented neural network

As observed in the previous section, mitigating the ill-conditioning of the Jacobian matrix is crucial to enabling PINN to solve complex problems. In fact, the controlled system not only provides a visualization of the ill-conditioning in PINNs but also presents a solution. The key challenge lies in how to substitute known quantities for $q_s$ in the controlled system, as it is always not available in practice. A simple strategy is to substitute the output $q_n = \hat{q}(\cdot; \theta_n)$ of the neural network at the $n$th optimization step for $q_s$, yielding

$$f(q) - \gamma(q - q_n) = 0 \qquad (8)$$

We minimize the residual of Equation (8), referred to as inner iteration. When it achieves appropriate convergence after $K$-step optimization, we proceed to the next optimization step, and update $q_n$ with the latest network output, referred to as outer iteration. Algorithm 1 provides detailed steps. In practice, the inner iteration in Algorithm 1 can be replaced by the internal iterations of the L-BFGS algorithm, where the maximum number of iterations is set to *K*. For the L-BFGS optimizer, we restart it at each outer iteration to adapt to changes of the loss function, while resampling collection points randomly. In the event of encountering *NaN* (Not a Number) or other failures during model training, we discard the current model and load the model of the previous outer iteration step to continue training. Based on our experience, these strategies greatly enhance the robustness of the training. Furthermore, these strategies also address the incompatibility between the L-BFGS optimizer and batching, eliminating the necessity to provide a dataset representing the full solution at each epoch. This enables the solving of large-scale problems, such as high-dimensional parametric problems [17].

**Algorithm 1:** TSONN
**Input:** Initial $\boldsymbol{\theta}$, collocation points, outer iterations *N*, inner iterations *K*, gain $\gamma$.
1: for $n = 1, 2, \cdots, N$ do
2: $\quad \boldsymbol{q}_n = \boldsymbol{q}(\cdot; \boldsymbol{\theta})$
3: $\quad$ for $k = 1, 2, \cdots, K$ do
4: $\quad\quad$ (a) Compute the loss $\mathcal{L}(\boldsymbol{\theta}) = \frac{1}{N} \| \boldsymbol{f}(\boldsymbol{q}(\cdot; \boldsymbol{\theta})) - \gamma(\boldsymbol{q}(\cdot; \boldsymbol{\theta}) - \boldsymbol{q}_n) \|^2$
5: $\quad\quad$ (b) Update the parameters $\boldsymbol{\theta}$ via gradient descent $\boldsymbol{\theta} \leftarrow \boldsymbol{\theta} - \eta \nabla_{\boldsymbol{\theta}} \mathcal{L}(\boldsymbol{\theta})$
6: $\quad$ end
7: end
**Output:** $\hat{q}(\cdot; \boldsymbol{\theta})$

Despite different starting points, Algorithm 1 is partially similar to the time-stepping-oriented neural network (TSONN) proposed in our previous study [26], which introduces pseudo-time derivative to decompose the originally ill-conditioned problem of PINNs into a series of well-conditioned sub-problems over given pseudo time intervals. Indeed, Equation (8) can be readily transformed into the following implicit pseudo time-stepping scheme, where the gain $\gamma$ is equivalent to $1/\Delta\tau$.

$$\frac{q - q_n}{\Delta\tau} = f(q) \tag{9}$$

We refer to it as pseudo time-stepping rather than time-stepping for two reasons. The first reason is that the current algorithm is applicable to both time-independent and time-dependent problems, where the latter already involve the presence of a time dimension. The second reason is that we are not concerned with whether the convergence history of TSONN accurately follows the temporal evolution process, but rather focus solely on the accuracy of the converged solution.

To emphasize the strong correlation of Algorithm 1 with time-stepping, we still refer to it TSONN. This current work can be regarded as an explanation and refinement of TSONN. An important improvement is that, based on the current analysis, linear forcing terms (or pseudo-time derivative terms in TSONN) should not only be applied to the PDE operator but also to the boundary condition operator and the initial condition operator (if any). To expedite convergence, for the initial condition operator or the Dirichlet boundary condition operator, known values can be directly substituted for $q_s$ instead of using the predicted values of the neural network. In this case, it is equivalent to the relative weights varying with gain. When Algorithm 1 is associated with time-stepping, its global convergence is ensured by one of the main principles of dynamics, namely that stable dynamical systems converge to their steady solutions for any initial value after a sufficiently long period of time. In addition, negative gain implies a "reversed-time" system, which is ill-posed as it is impossible to deduce the flow history from its steady-state solution.

Note that having the correct form for $f(q)$ is crucial for TSONN. In PINNs, Equation (5) can be expressed in four equivalent forms, as listed in Equation (10). However, their Jacobian matrices are different, and only one of these forms can ensure that the system has entirely negative eigenvalues. A simple guideline is to ensure that the Jacobian matrices of the PDE operator, boundary condition operator, and initial condition operator (if any) each have negative eigenvalues. Some typical examples are, for the form of $f_1(q)$, the Burgers' equation should be $g(q) = -q_t - qq_x + \nu q_{xx}$, the Poisson' equation should be $g(q) = q_{xx} + q_{yy}$, and the initial condition and Dirichlet boundary condition operators should be $h(q) = c - q$, where $c$ is a known target value.

$$f_1(q) = \begin{bmatrix} \lambda_{PDE} g(q) \\ \lambda_{BC} \sqrt{N_g / N_h} \, h(q) \end{bmatrix}, f_2(q) = \begin{bmatrix} -\lambda_{PDE} g(q) \\ -\lambda_{BC} \sqrt{N_g / N_h} \, h(q) \end{bmatrix},$$
$$f_3(q) = \begin{bmatrix} -\lambda_{PDE} g(q) \\ \lambda_{BC} \sqrt{N_g / N_h} \, h(q) \end{bmatrix}, f_4(q) = \begin{bmatrix} \lambda_{PDE} g(q) \\ -\lambda_{BC} \sqrt{N_g / N_h} \, h(q) \end{bmatrix}$$
(10)

The original version of TSONN has been demonstrated to successfully solve the Allen-Cahn equation (time-dependent system) and the lid-driven cavity flow within the Reynolds number range of 100 to 5,000 (time-independent system). In this paper, we consider complex engineering problems governed by the incompressible Navier-Stokes equation.

4.2 The flow around the NACA0012 airfoil

We first consider the flow around the classical NACA0012 airfoil with Re = 5,000. Despite TSONN inherits the mesh-free property of PINNs, we still employs the mesh points from a background grid (Figure 4) as collocation points. The mesh distribution offers fundamental insights into boundary layers. The wall boundary conditions are $u=0, v=0$. The velocity inlet boundary conditions are $u=\cos(\alpha), v=\sin(\alpha)$, where $\alpha$ is the angle of attack. The pressure outlet boundary conditions is $p=0$. We adopt a volume-weighted PDE residual, which has been validated to perform better in non-uniform mesh distributions and has successfully solved airfoil flow at Re = 400 [20]. Consequently, $g(q)$ is redefined as:

$$g(q) = \frac{s}{\|s\|_2 / \sqrt{N_g}} \tag{11}$$

where the vector $s$ encompasses the grid volume at each residual point, and the denominator in Equation (11) normalizes the influence of $s$. It is important to note that both volume weighting and relative weighting alter the condition number of the underlying Jacobian matrix. Therefore, as a general strategy for mitigating ill-conditioning, pseudo-time stepping should be positioned at the outermost layer of the loss function. This is also the fundamental reason for our reformulation of the PINNs loss as Equation (4). We solve this system with varying pseudo time steps $\Delta \tau$ using a network with 5 hidden layers and 128 neurons per hidden layer. The relative weights are $\lambda_{PDE} = 100$ and $\lambda_{BC} = 1$.

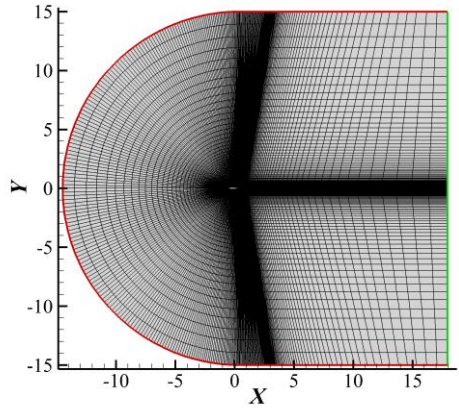

Figure 4. Background grid of the NACA0012 airfoil. The red boundary represents the velocity inlet, while the green boundary represents the pressure outlet.

Figure 5 (a) and (b) present the convergence histories of the relative $L_2$ error for different pseudo time steps. We observe that PINNs fail to achieve any meaningful result, highlighting its ill-conditioning. When the time step size is large ($\Delta \tau = 1/\gamma = 10$), TSONN obtains convergence histories almost identical to PINNs, validating PINNs as a special case of TSONN when the pseudo time step tends to infinity (i.e., gain tends to 0). As the time pseudo step decreases (i.e., the gain increases), the PDE system becomes increasingly well-conditioned, thus achieving more stable convergence. In the controlled system of Figure 1, the introduction of steady-state solutions leads to faster convergence with smaller pseudo time steps. However, for TSONN, smaller pseudo time step implies more iterations required for the flow to evolve from an initial value to its steady-state solution, potentially resulting in slower convergence, as depicted in Figure 5. In addition, smaller pseudo time steps typically entail more pronounced oscillations in convergence histories, consistent with CFD experience. Ultimately, TSONN achieves a relative error of 2%. The pressure contour plots and wall pressure coefficient distributions are illustrated in Figure 6 and Figure 7, respectively.

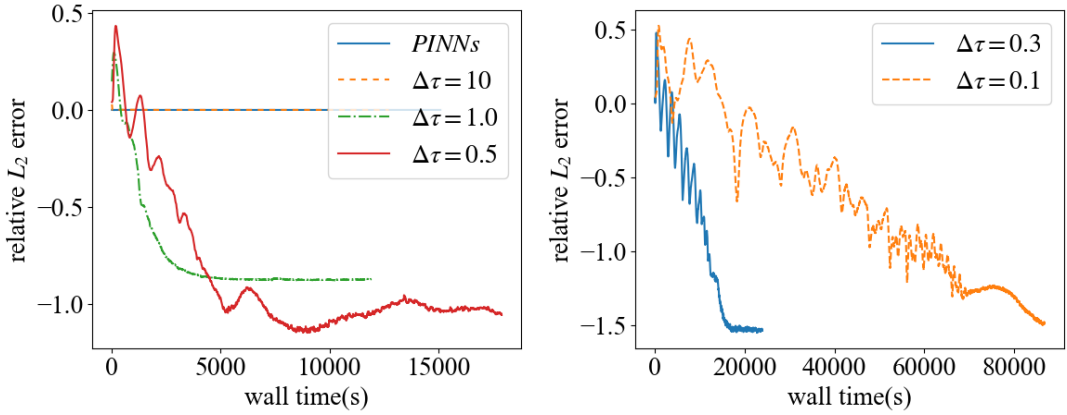

Figure 5. The convergence history of the relative L$_2$ error obtained by PINNs and TSONN with different pseudo time steps.

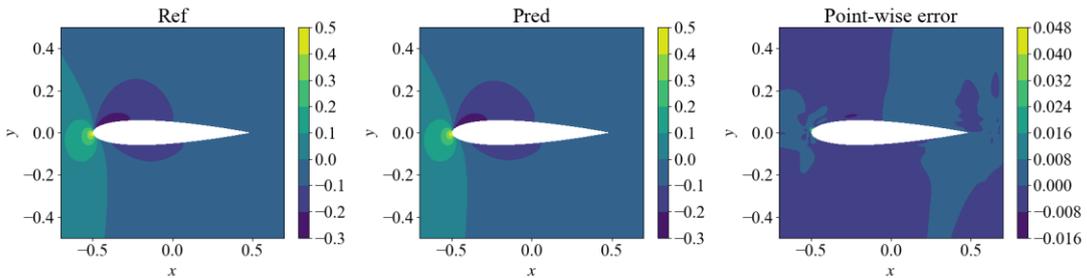

Figure 6. The pressure field near the wall obtained by TSONN with a pseudo-time step of 0.3.

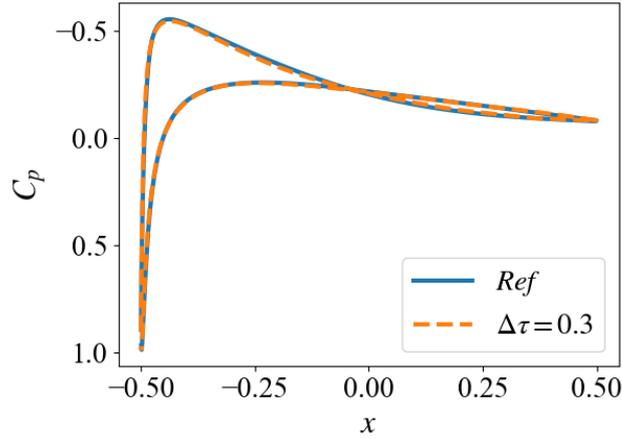

Figure 7. The wall pressure coefficient distribution obtained by TSONN.

4.3 The flow around the M6 wing

We next consider a complex three-dimensional problem, the flow around the M6 wing, which is a benchmark geometry in CFD. We consider Re = 5,000, which is nearly the maximum Re permitted for laminar flow. Higher Re necessitates coupling the Navier-Stokes equations with turbulence models, which will be considered in future studies. Figure 8 depicts the background grid. The boundary conditions are similar to those of the two-dimensional NACA0012 airfoil. We increase the hidden layers of the network from 5 to 8 to enhance the representation capability of the neural network. According to the author's experience, similar types of problems can often share the same relative weights [27]; therefore, the relative weights are still set as $\lambda_{PDE} = 100, \lambda_{BC} = 1$. We choose $\Delta\tau = 0.3$.

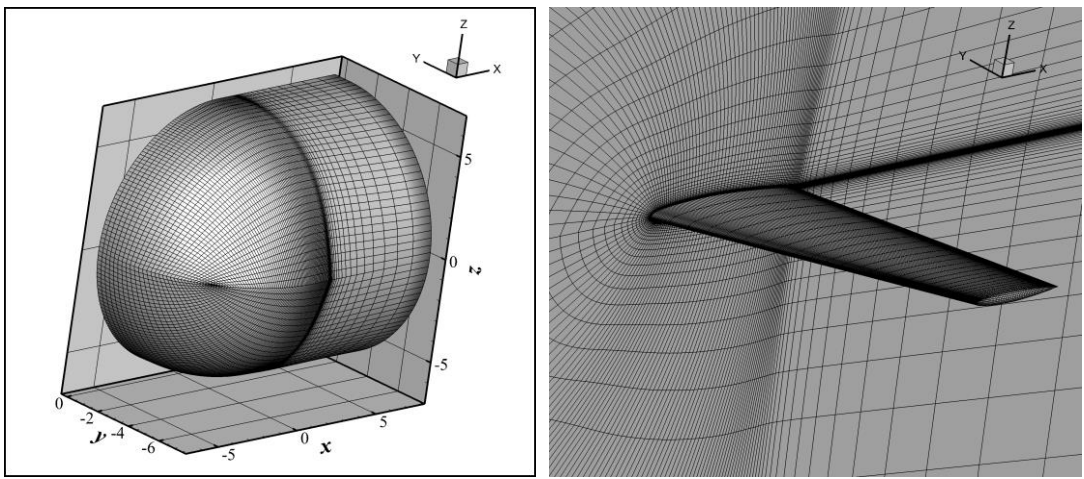

Figure 8. Half of the Background grid of the M6 wing.

As depicted in Figure 9, we have successfully resolved the flow, yielding a relative error of 6% in the wall pressure distribution. We observe a larger error occurring at the leading edge of the wing root, coinciding with a local sharper transition in flow. To facilitate clearer comparison, Figure 10 shows *x-p* line plots at different y-coordinates. Apart from a significant discrepancy at the leading edge of the wing root (i.e. y=0), the results obtained from TSONN closely match with the reference solution at other locations.

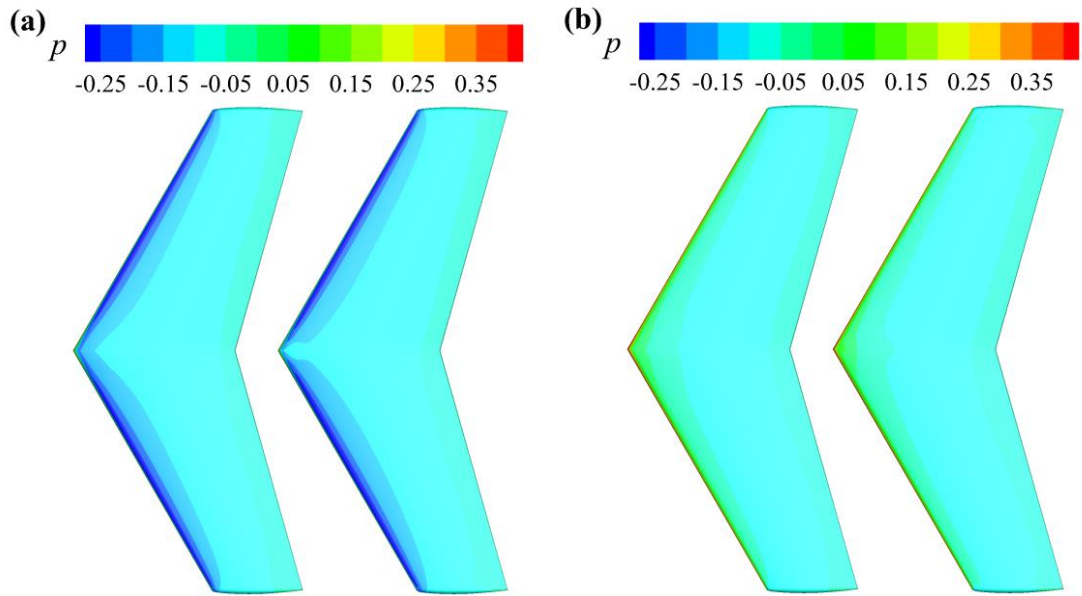

Figure 9. The wall pressure distribution of (a) the upper surface and (b) the lower surface obtained by TSONN. Left: Ref. Right: TSONN.

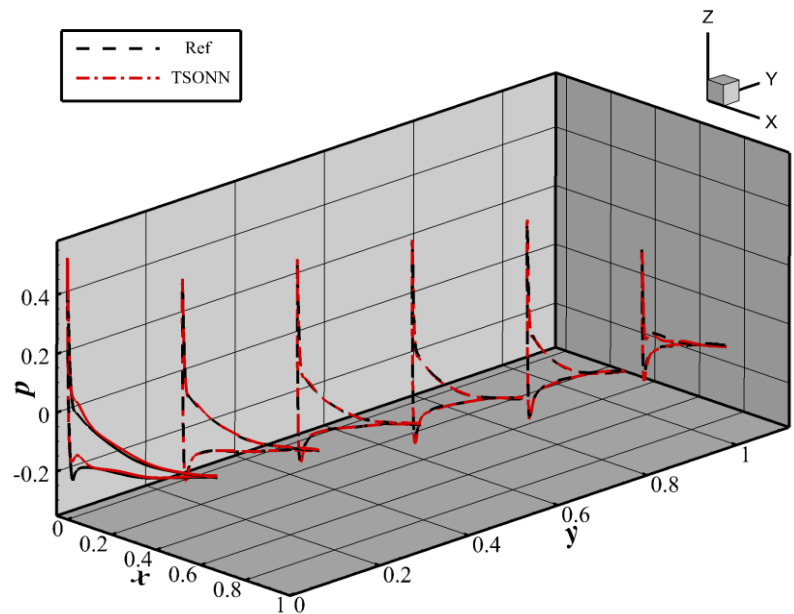

Figure 10. The wall pressure coefficient distribution at different y-coordinates.

# 5 Conclusions

This study demonstrates a physical perspective for understanding the ill-conditioning of PINNs. We construct a controlled system with an adjustable condition number of the Jacobian matrix, validating that the ill-conditioning of PINNs primarily originates from the ill-conditioning of the Jacobian matrix. As the condition number of the Jacobian matrix decreases, PINNs achieve faster convergence and higher accuracy. Therefore, we suggest that mitigating the ill-conditioning of PINNs should focus on alleviating the ill-conditioning of the Jacobian matrix, although the beneficial contributions of machine learning techniques such as neural network architectures and optimization algorithms to PINNs should not be overlooked.

Building upon this understanding and the natural extension of controlled systems, we propose an approach that transforms the ill-conditioned optimization problem of PINNs into a sequence of well-conditioned sub-optimization problems. Physically, this algorithm can be interpreted as aligning the convergence process of the neural network with the (pseudo) temporal evolution of the physical system, hence refereed as time-stepping-oriented neural network (TSONN). An important feature of this approach is that it retains almost all advantages of PINNs. We successfully solves the flow around the M6 wing at a Reynolds number of 5,000. Our analysis and results suggest that PINNs represent a special case of TSONN when the time step tends to infinity. Smaller time steps in TSONN lead to more well-conditioned sub-optimization problems but may necessitate more steps to converge to steady solutions.

In future research, an adaptive time step or local time step for TSONN may be warranted. Additionally, we recommend exploring preconditioning techniques to alleviate the ill-conditioning of PINNs, potentially achieving more efficient convergence.

## Data Availability Statement

The data that support the findings of this study are available from the corresponding author upon reasonable request.

## Conflict of Interest Statement

The authors have no conflicts to disclose.

## Acknowledgments

We would like to acknowledge the support of the National Natural Science Foundation of China (No. 92152301).